# TexComp - A Text Complexity Analyzer for Student Texts

*Tuomo Kakkonen*

University of Joensuu, Finland

**Abstract:**

*This paper describes a method for providing feedback about the degree of complexity that is present in particular texts. Both the method and the software tool called TexComp are designed for use during the assessment of student compositions (such as essays and theses). The method is based on a cautious approach to the application of readability and lexical diversity formulas for reasons that are analyzed in detail in this paper. We evaluated the tool by using USE and BAWE, two corpora of texts that originate from students who use English as a medium of instruction.*

## 1  Introduction

Text complexity is one of the components in the definition of *style* in any written composition. *Lexical diversity* (LD) is a term used to indicate the range of the vocabulary that is present in a text. Lexically diverse text is usually regarded as being more competent and more persuasive in its effect than an equivalent low-diversity reproduction of the same text. LD measures have been applied, for example, in dating literary works, in the creation of a set of stylistic "fingerprints" that enable to identify the author(s), and for assessing the overall quality of a text. *Readability* refers to the ease of reading, especially as the result of the writing style. Formulas for determining the readability of texts have been applied in several fields such as the enforcement of the "plain language laws" for insurance policies and in the practice of journalism wherever there is a need to simplify newspaper texts.

In educational context, the automatic identification of text complexity can be used, for example, as a basis for offering automatic feedback as well as a tool for guiding human assessors. One of the possible uses is to detect "bad faith essays". This term is used to refer to essays that have been deliberately constructed in order to deceive an automatic essay grading system into giving the student concerned a high grade.

This paper is organized as follows: Section 2 reviews the background of this work. Sections 3 and 4 respectively describe our new method and provide details about our evaluation of the TexComp tool. The paper closes with Section 5 which summarizes our findings.

## 2  Previous Work

### 2.1  Lexical diversity and readability measures

The rationale for making LD measurements is to create a numerical quantification that indicates the linguistic complexity of that particular text [2]. Most of the existing LD measures are based on the *word type-token ratio* (TTR) (the number of word types (i.e. different words) divided by the number of word tokens). The main problem associated with such LD formulas is their tendency to produce inexact quantifications when texts of varying lengths are being compared [2, 4]. In order to be truly useful, a method of measuring the diversity of vocabulary has to be independent of text length after a certain minimum threshold for length has been exceeded.

As Fry [1] points out, over one hundred readability formulas have been described in the scientific literature. The number of words and sentences in the text as well as the proportion of "difficult" words are the most common text statistics that these formulas utilize. The use of readability formulas are often criticized (see, for example, [5, 6]). The main of criticism devolves on the fear that they may encourage to "write to the formula", to compose texts in such a way that the end product will maximize the readability score. Klare [7], however, points out the following advantages of readability formulas:
- A more readable text permits an increase in the reading speed of the average reader.
- While a high readability score does not necessarily mean that the text is easily comprehensible, such a statement is true in a sufficient number of cases to make the effort of producing readable texts a worthwhile undertaking.

Most of the criticism of readability formulas appears to be directed against the *improper use* rather than the whole concept of a formula itself. Many researchers such as Klare [7] and Redish [5] agree that if these formulas are *carefully* used, they can function as useful tools in the assessment of writing.

### *2.2 Stylistic assessment of student texts*

One of the first systems to provide automatic feedback on student texts was Writer's Workbench [8]. The guiding design principle behind the system was that the problem of poor student writing could be ameliorated if "style guides were automated". Among the stylistic features that the system utilized were the average length of the words and sentences, and "several readability indexes". Some automatic essay grading systems use readability scores [3, 9]. Since most of these systems are commercial products, no exact information has been released about the way in which these formulas are applied. It is, however, evident that these grading systems use stylistic calculations in two ways: firstly, as a component for defining a holistic grade for an essay, or, secondly, for providing a trait grade for writing style.

## 3 TexComp

TexComp takes precautions against known pitfalls in applying readability and LD measures. Firstly, the exact values of the readability and LD formulas are never shown to the student or the teacher as they are. Rather, the data provided by the formulas is used for indicating exceptional texts. Secondly, in order to minimize the effect produced by possible consistency problems with a single readability or LD formula, TexComp uses averages from two formulas to determine the readability and lexical complexity of a passage of text. Thirdly, TexComp's calibrated mode allows to set the threshold values for feedback on the basis of training data accumulated from other students' texts, rather than being forced to rely on controversial grade level definitions used in many readability formulas.

### 3.1  Lexical diversity measures

The criteria for selecting the LD measures was based on the need to achieve a high degree of language-independency and accuracy. McCarthy's [4] dissertation was particularly useful for selecting the two LD measures, namely Yule's K [10] and Maas's $a^2$ [11]. Both these measures ranked among the best in McCarthy's empirical evaluation, specifically with regard to their length-independence. Another distinct advantage is that both these measures remain relatively straightforward to understand, implement and modify for different languages. Yule's K measures the rate at which words are repeated in a text. Many variations of Yule's K formula are available. We use the definition from [2]. A high *K* value means that the vocabulary is concentrated and words are repeated over and over again. A small *K* value, by contrast, indicates that the vocabulary is more varied and less concentrated onto a few special words. As with Yule's K, smaller $a^2$ values denote greater LD.

The TexComp LD score, *TCLD*, is defined as the weighted average $(2K+a^2)/2$. K is accorded a weight of two in order to prevent too much emphasis being given to Maas' $a^2$ measure, which tends to be approximately double of the value of K for any given text. It is important to note TexComp's cautious approach in the use of LD measures. No claim is being made that LD could be used as an essential measure of the overall quality of a text or for predicting the general language proficiency of a writer [12].

### 3.2  Readability measures

The criterion of language-independency guided our selection of readability measures to *läsbarhetsindex* (LIX) [13] and its modification RIX [14]. Another reason for choosing these particular measures was that, in contrast to many other readability formulas, they do not require syllable counting, which is potentially inaccurate and also makes it more difficult to modify a readability formula to multiple languages. In contrast to some other measures, LIX and RIX are also based on observation of the whole text rather than merely on sampling. Higher LIX and RIX scores indicate more complex texts. The overall TexComp readability score (*TCR)* of a text is defined as (10*RIX + LIX)/2. The RIX value is scaled in order not to overemphasize LIX because the LIX value for the same text tends to be approximately ten times larger than the RIX value for the same text.

Since we took due heed of the warnings about the use of readability formulas described in Section 2.2, TexComp exemplifies a cautious approach to the use of these measures. TextComp thus uses formulas only to assess text complexity and makes no claim that readability formulas can measure any of the other properties of text. And the system never reveals readability values either to the teacher or to the student, but utilizes them as the basis for feedback.

### 3.3  Uncalibrated and calibrated modes

TexComp provides feedback to the user by comparing the *TCLD* and *TCR* values to the threshold values $TCLD_{min}$, $TCLD_{max}$, $TCR_{min}$ and $TCR_{max}$ These latter values are used to determine whether the document is below or above the limits of "normal" readability and LD. This information can be used, among other things, to alert a teacher to a text that manifests a sub-standard use of vocabulary or to obtain positive feedback about a highly readable text. TextSylist, for example, will report that a text utilizes an overly simple vocabulary if the TCLD value is greater than $TCLD_{max}$. TexComp has two modes of operation: (1) the *uncalibrated mode*, in which the system comments on LD and readability on the basis of preset threshold values that remain unchanged for different text collections if the user does not change them. Table 1summarizes the default values that were set during the system development phase based on empirical experiments.

**Table 1.**The default feedback threshold values.

| Measure | Threshold | |
|---|---|---|
| | *Min* | *Max* |
| *TCLD* | *150* | 250 |
| *TCR* | *40* | 80 |

It is obvious that it is inadmissible to use the same criteria to assess a text written by high school pupil and a text written by a postgraduate student at a university. The *calibrated mode* can be used for document collections which contain a sufficient number of documents on the same topic to make it possible to set threshold values. The rationale behind calibration is that in order to be appropriate, a student's text should not differ drastically in terms of complexity (readability and LD) from texts produced by other students who are taking the same course or using the same learning materials prescribed for the course. Lack of space unfortunately prevents us from supplying further details of how the calibrated mode functions.

# 4  Evaluation

## 4.1  Test data and settings

We selected the following two corpora of written student texts in English for the evaluation study: the *British Academic Written English* (BAWE) corpus [15] and the *Uppsala Student English* (USE) corpus [16]. These two corpora are the only collections of student texts that we are aware of that contain information about the study level of the writers of the texts. The USE corpus consists of 1,490 texts written by 440 non-native English-speaking students in the Department of English, Uppsala University, Sweden. We divided the corpus into two parts: those that had been written by first-term students (T1) and those that had been written by second-term students of English (T2). The BAWE corpus consists of student texts collected from the following three universities: Oxford Brookes, Reading and Warwick University. These texts were written by 1,039 native or partly nativized English-speaking students in each of three undergraduate years and a selection who were engaged in postgraduate studies. Table 2 offers all the details of the test data that we used in the experiment.

**Table 2**. The test data

| Corpus | Subcorpora | | | | Total |
|---|---|---|---|---|---|
| *USE* | *T1* | | *T2* | | 1,490 |
| | 1,238 | | 252 | | |
| *BAWE* | *1* | *2* | *3* | *4* | 2,752 |
| | 807 | 767 | 591 | 587 | |

Our first hypothesis was that in order to prove that the readability and LD values were reliable in their measurement of the differences in writing style between non-native and native students, there should be a noticeable difference between the average *TCR* and *TCLD* values assigned for our two test corpora. Our second hypothesis was that in order to prove that TexComp is reliable in the way it makes distinctions between the writing styles of students at different levels of study, the average *TCR* and *TCLD* values should also be different for each of the six subcorpora. Our assumption was that the *TCR* values would increase and the *TCLD* values would decrease according to (1) whether the students concerned were a native or non-native speakers of English and (2) the study level presented by each subcorpus. Section 4.2.1 reports on these experiments.

We also observed the proportion of texts in each subcorpus that were identified as exceptional according to the uncalibrated assessment mode (Section 4.2.2). Our hypothesis was that as the students' academic level increased, the proportion of texts that were bellow the negative feedback threshold would decrease and the proportion of texts that were above the positive feedback threshold would increase. Finally, we undertook a qualitative sampling study of some of the documents that had been identified as exceptional (Section 4.2.3).

## *4.2  Results*

### 4.2.1   Detecting differences between subcorpora

As the results in Table 3 show, the average *TCLD* and *TCR* values changed between the non-native (USE) and native (BAWE) corpora by -13.8% and +45.6% respectively. This supports our hypotheses that both LD and the readability of texts written by native speakers should be better than the same factors in texts written by non-native speakers. Figure 1 plots the *TCLD* and *TCR* values for individual subcorpora.

**Table 3**. The test data. Columns TCLD and TCR show the average values for each subcorpus.

| Corpus | Subcorpus | Results | |
|---|---|---|---|
| | | *TCLD* | *TCR* |
| USE | *T1* | 212.6 | 41.4 |
| | *T2* | 202.5 | 45.3 |
| | *AVG* | 210.8 | 42.1 |
| BAWE | *1* | 188.8 | 58.7 |
| | *2* | 181.8 | 62.2 |
| | *3* | 182.1 | 62.2 |
| | *4* | 174.1 | 63.3 |
| | *AVG* | 181.7 | 61.3 |

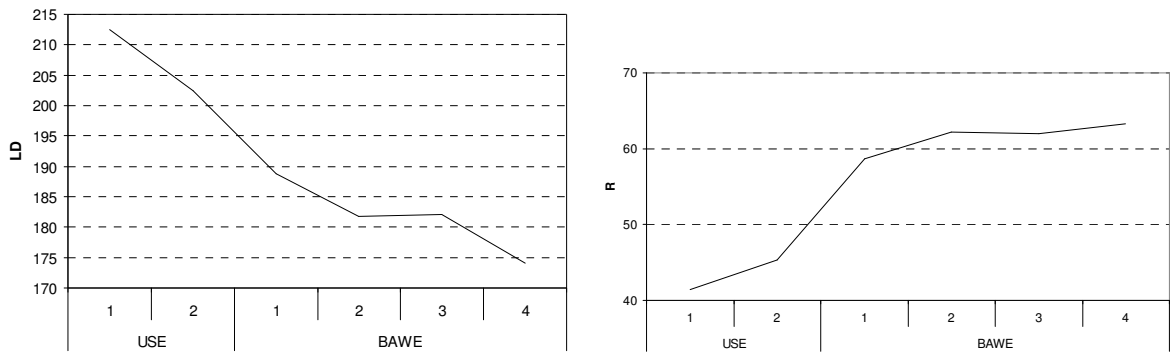

**Figure 1**. Lexical diversity and readability values.

Figure 1 show that our hypotheses of falling *TCLD* values and rising *TCR* values holds true for all the other subcropora expect for those of the third year students in the BAWE corpus. Apart from the most obvious explanation that the writing style of the students had undergone no development during the last two years of the undergraduate studies of these students, there are other possible explanations for this phenomenon. Apart from variations in the size of different subcorpora, the variation in the writer population and text genre may be contributing factors; The texts in different subcorpora were not about the same topics, nor were they in all cases written by the same students. It is interesting to note that this same phenomenon occurs with regard to both readability and LD assessment. This indicates that the two models are consistent with one another. At the same time, most of the documents that were categorized as exceptional by one of the measures were not identified as exceptional by the other measure. This proves that the *TCR* and *TCLD* values do not measure the same property and that they can therefore be used to complement one another for the process of generating feedback on text complexity.

### 4.2.2 Threshold values for feedback

In order to verify that the uncalibrated threshold values for providing feedback behave in the expected way, we repeated the experiment in the way reported in subsection 4.2.1 and observed the proportion of documents that received *TCLD* and *TCR* values that were not within the two thresholds. Our hypotheses was that TexComp would find smaller and smaller proportions of documents with negative complexity characters as the English language skills and academic level of the students increased from one subcorpus to another. Table 4 summarizes the results of this experiment.

**Table 4**. Percentages of texts that received either higher than the maximum (H) or lower than the minimum (L) threshold value for *TCLD* and *TCR*

| Corpus | Sub-corpus | TCLD | | TCR | |
|---|---|---|---|---|---|
| | | *L* | *H* | *L* | *H* |
| USE | *T1* | 5.4 | 0.0 | 46.2 | 0.1 |
| | *T2* | 3.2 | 0.8 | 29.4 | 0.0 |
| | *AVG* | 5.0 | 0.1 | 43.3 | 0.1 |
| BAWE | *1* | 2.9 | 3.7 | 5.0 | 7.4 |
| | *2* | 0.7 | 4.8 | 2.0 | 8.2 |
| | *3* | 1.7 | 3.9 | 2.7 | 9.8 |
| | *4* | 0.7 | 9.9 | 1.4 | 12.4 |
| | *AVG* | 1.7 | 5.1 | 2.9 | 9.2 |

As the results in Table 4 show, a higher proportion of the texts in the USE corpus were deemed to have low readability and LD characteristics than the texts in the BAWE corpus. In contrast to the USE documents, among which only a few received readability and LD scores that were better than the threshold values for a high-quality text, 5.1 and 9.2 per cent of the BAWE documents scored better than the thresholds $TCLD_{min}$ and $TCR_{max}$. As a result of this, we were able to confirm that our hypothesis that the proportions of documents that exceeded or were lower than the corresponding threshold values (i.e. that were identified as having negative and positive complexity features) was correct, with the exception of the BAWE 2 subcorpus. The high proportion of documents in the USE corpus that were diagnosed as having readability issues (43.3%) exemplifies the need for the calibrated mode or for being able manually to adjust the threshold values to reflect the abilities of different student groups.

### 4.2.3  Samples

A closer look at the results supports the usefulness of the method. For example, the sentence from a document in the BAWE corpus that received the highest *TCR* value in the whole experiment (namely 126.3), clearly indicates a good writing style: "*Proponents of this change have argued that traditional, public-funded, on-station research does not address the needs and problems of poor farmers in 'complex, diverse and risk-prone' areas, because their physical and socioeconomic conditions are too different from those at the research stations*". The sentence used as an example below is from a document in the non-native USE corpus. This document was assigned a high *TCLD* value, and the following sentence clearly demonstrates the problems that the writer was having with word choices: "*It's difficult to find the right words in my head and to get them down on paper (or on to the screen).*"

### *4.3  Discussion*

The evaluations reported above have shown that the two text complexity measures, *TCR* and *TCLD*, can be used for indentifying levels of complexity in student texts and that this information can then be used to detect those texts that need improvement and to provide positive feedback on high-quality texts. The main limitations of the current work concern the evaluation scheme and language-independency of the devised methods. LIX and RIX scores were originally developed for Swedish, and the set limit for long words (>6 characters) reflects the properties of that particular language. However, studies (such as, for example, [1]), show that the two measures work well in a variety of languages such as English, French, German and Greek. Swedish and English are, moreover, both Germanic languages with very limited inflectional systems. This supports the idea that the same method of detecting text complexity could well be applied to both these languages. The application of the methods presented in this paper to languages with rich morphological systems (such as Finnish or Turkish) calls for the use of a morphological analyzer to identify the base forms of each word. The cut-off points for long words would, moreover, need to be changed to reflect the properties of the language that is being analyzed.

Because of the lack of corpora in which the writing style, readability or LD have been manually marked up, we had to devise the evaluation scheme that is reported above. While our evaluation scheme may be less than perfect, it supports our basic claim that the *TCLD* and *TCR* values reflect the writing style of students on the basis of complexity measures and that this information can be used for identifying texts with the characteristics of high and low text complexity.

## 5  Conclusion

We have described a method for providing feedback on student texts on the basis of text complexity measures. The TexComp tool that implements the proposed method can be used for detecting stylistically divergent texts from a set of student texts. This paper also described the possible limitations of the measures that the system applies, and offered reasons for the various decisions that we made when designing the method and the TexComp tool. Our evaluation using two freely available corpora of student texts supported our basic argument that the tool is able to measure differences between texts that have been written by students who vary in their mastery of English and their level of academic experience. This permits the user to detect outliers from a set of student texts by using two methods – the calibrated and the uncalibrated. Future work possibilities for further developing text complexity measurement methods and the TexComp tool include creating evaluation materials that have been marked up manually with stylistic information. Data of this kind would be needed to verify the effectiveness of the system in providing feedback that goes beyond the holistic assessment of the whole text.

## Author(s):

Tuomo Kakkonen, PhD
University of Joensuu, Department of Computer Science and Statistics
Department of Computer Science and Statistics, University of Joensuu
PO Box 111, 80101 Joensuu, Finland
tuomo.kakkonen@cs.joensuu.fi